\documentclass[reprint,superscriptaddress,amsmath,amssymb,aps,prb,floatfix]{revtex4-1}

\usepackage{graphicx}
\usepackage{color}
\usepackage[subnum]{cases}
\usepackage{soul}
\usepackage{verbatim}
\renewcommand{\v}[1]{{{\bf #1}}}

\newcommand{\dy}[1]{{\overset{\text{\tiny$\leftrightarrow$}}{\rm #1}}}
\newcommand{\ay}[1]{{\bar{\bf #1}}}

\newcommand{\mx}[1]{{\bar{\bar #1}}}
\newcommand{\rfig}[1]{Fig.~\ref{fig:#1}}
\newcommand{\reqn}[1]{Eq.~\eqref{eq:#1}}

\def\Xint#1{\mathchoice
   {\XXint\displaystyle\textstyle{#1}}
   {\XXint\textstyle\scriptstyle{#1}}
   {\XXint\scriptstyle\scriptscriptstyle{#1}}
   {\XXint\scriptscriptstyle\scriptscriptstyle{#1}}
   \!\int}
\def\XXint#1#2#3{{\setbox0=\hbox{$#1{#2#3}{\int}$}
     \vcenter{\hbox{$#2#3$}}\kern-.5\wd0}}
\def\dashint{\Xint-}

\begin{document}


\title{Clustering Diffused-Particle Method for Scattering from Large Ensembles of Electromagnetically Polarizable Particles}

\author{Lang Wang}
\affiliation{Department of Electrical \& Computer Engineering, University of Illinois at Urbana-Champaign, Urbana, IL 61801, USA}
\affiliation{The Institute of Optics, University of Rochester, Rochester, NY 14627, USA}
\author{Ilia L. Rasskazov}
\author{P. Scott Carney}
\email{Corresponding author: scott.carney@rochester.edu}
\affiliation{The Institute of Optics, University of Rochester, Rochester, NY 14627, USA}

\begin{abstract}
The Foldy-Lax equation is generalized for a medium which consists of particles with both electric and magnetic responses.
The result is used to compute fields scattered from ensembles of particles.
The computational complexity is reduced by hierarchical clustering techniques to enable simulations with on the order of $10^{10}$ particles.
With so many particles we are able to see the transition to bulk media behavior of the fields.
For non-magnetic materials, the observable index, permittivity, and permeability of the effective bulk medium are in good agreement with the Clausius-Mossotti relation.
The fields simulated for particles with both electric and magnetic responses are in good agreement with new analytical results for a generalized effective medium theory~\cite{Lang_Ilia_Scott_1}.
\end{abstract}

\maketitle

\section{Introduction}

Light scattered from large collections of atoms, molecules and particles appears to propagate according to the macroscopic Maxwell's equations with permeability and permittivity that emerge from the microscopic components of the constituents.
Solutions to the scattering problem at the macroscopic scale are important to understand how light interacts with matter and to engineer optical systems.
Self-consistent solutions to the scattering problem at the microscopic scale are important in understanding strongly-interacting systems but present challenges of computational complexity.
Moreover, techniques developed to date have not provided a means to include particles which have both electric and magnetic responses simultaneously.
Here we present a multiscale method that allows us to simulate the scattering of light from collections of point-particles with numbers of the order of $10^{10}$ using modern computer hardware.
Moreover, we provide a means for self-consistent solutions for particles which respond both to electric and magnetic fields.

The applications of the results presented here include design of nanostructures~\cite{Kelly2003TheEnvironment,Jain2006CalculatedBiomedicine,Jain2006PlasmonModel,perez-juste_gold_2005,Jin2003ControllingExcitation,Huang2009GoldApplications,Chen2005GoldAgents,Myroshnychenko2008ModellingNanoparticles,Hartland2011OpticalNanostructures,Rasskazov2016OvercomingChains}, nanosensing~\cite{Jain2008NobleNanosensing,giannini_plasmonic_2011}, localized surface plasmon resonance spectroscopy~\cite{Willets2007LocalizedSensing,Stiles2008Surface-EnhancedSpectroscopy,Ghosh2007InterparticleApplications,Lee2006GoldComposition,Noguez2007SurfaceEnvironment,Jain2007OnEquation,Klar1998Surface-plasmonNanoparticles,wiley_maneuvering_2006}, surface-enhanced Raman spectroscopy~\cite{Tian2002Surface-enhancedNanostructures,vanDijk2013CompetitionSpectroscopy}, atmospheric science~\cite{Bond2013BoundingAssessment,Dubovik2006ApplicationDust,Dubovik2000AccuracyMeasurements,Watson2002Visibility:Regulation}, and astronomy~\cite{Weingartner2001DustCloud}.
Several numerical techniques have been developed to compute the electromagnetic field scattered from large collections of particles or objects.
One of the most efficient and widely used tools is the T-matrix approach~\cite{Waterman1971SymmetryScattering,varadan_acoustic_1980,Egel2017a,Pattelli2018RoleSystems}, particularly suitable for particles with morphological complexity, with large sizes, or at resonance.
The discrete dipole method can simulate point dipoles interacting with one another via electric fields~\cite{Draine1993BeyondApproximation,Draine1994Discrete-dipoleCalculations}.
The application of fast algorithms and parallel computing has enabled the simulation of numbers of particles on the order of $10^8$ particles~\cite{Yurkin2011}, applying the approximation methods enables the simulation of numbers of particles on the order of $10^9$ particles~\cite{Penttila2021HowMedia}, whereas in the present work we consider $\sim 10^{10}$ particles.

In order to control the numerical complexity in our method, the particles are clustered, and those clusters are subsequently aggregated to form larger clusters~\cite{Song1997MLFMAObjects,Hackbusch1999A-Matrices,koc_multilevel_2001,jarvenpaa_broadband_2013,yurkin_discrete_2007,mulholland_light_1994}.
The scattered field from each element is calculated via the Foldy-Lax method~\cite{Foldy1945TheScatterers,Lax1951MultipleWaves,Lax1952MultipleSystems} for finding exact solutions of the field scattered from collections of point-particles.
In order to extend the method to particles with magnetic polarizability, we have found a generalization of the usual Foldy-Lax method.
Finally, we fit plane waves to the computed scattered fields to infer the macroscopic optical properties of the scatterers from the numerical results. 

The paper is organized as follows. 
In Sec.~\ref{theory} the usual Foldy-Lax approach is generalized to include particles with both electric and magnetic polarizabilities.
In Sec.~\ref{numerical}, we develop a hierarchical clustering method for solving the generalized Foldy-Lax method numerically. 
Finally, the new method is used to find the macroscopic properties of a large collection of particles, and future directions and applications are discussed. A SI system of units is used throughout the letter.

\section{Theory}
\label{theory}

\subsection{The Foldy-Lax equation}

Let us first consider the standard Foldy-Lax result~\cite{Foldy1945TheScatterers,Lax1951MultipleWaves,Lax1952MultipleSystems} for the scattering of an electric field from $N$ particles with the purely electric response. 
The electric field, $\v E_i$, on the $i$-th particle consists of the incident field and the field scattered by all the other particles:
\begin{equation}
\label{eq:Ed}
    \v E_{i}=\v E^{\rm inc}_{i}+\sum_{j\neq i}^{N} \v E^{\rm sca}_{ij}.
\end{equation}
Here $\v E^{\rm inc}_{i} = \v E^{\rm inc}(\v r_i)$ is the incident field, $\v r_i$ is the location of the $i$-th particle, and $\v E^{\rm sca}_{ij}$ is the electric field scattered by the $j$-th particle at location $\v r_j$ to the location $\v r_i$. 
Both the notations with subscript $i$ and with $(\v r_i)$ are used in this paper depending on the situation. 
A monochromatic field is considered without loss of generality and a time dependence of $e^{-i\omega t}$, where $\omega$ is the angular frequency of the field, is assumed and suppressed throughout the paper.

The particles polarized by the electric field are assumed to be point-like.
The dipole current is thus given by
\begin{equation}{\label{eq:I}}
\v J(\v r, \v r_{i})  = -i\omega\alpha_{\rm e0}\delta(\v r - \v r_{i})\v E(\v r_{i}).
\end{equation}
Here $\alpha_{\rm e0}$ denotes the complex-valued bare electric polarizability of a single particle, which includes the self-interaction~\cite{DeVries1998PointWaves}, $\v E^{\rm sca}_{ii}$.
The electric field scattered from a polarized particle is given by
\begin{equation}
\label{eq:E_sca}
\begin{split}
    \v E^{\rm sca}(\v r, \v r_{i}) = & \int i\omega\mu_{\rm 0} \dy G(\v r, \v r') \v J(\v r', \v r_{i}) {\rm d}^3 r'
    \\
    = & \omega^2\mu_{\rm 0}\dy G(\v r, \v r_{i}) \alpha_{\rm e0}\v E(\v r_{i}),\, \v r \neq \v r_{i}.
\end{split}
\end{equation}
Here $\mu_{\rm 0}$ is the magnetic constant, $\v r'$ is the location of the dipole, and $\v r$ is the observation location.
The dyadic Green's function $\dy G$ in free-space satisfies the equation
\begin{equation}
\label{eq:Green}
    [\nabla\times\nabla\times - \, k_{\rm 0}^2]\dy G(\v r, \v r')=\delta(\v r-\v r') \bar{\bar I}_{3}.
\end{equation}
Here $\nabla\times$ denotes the curl operator acting on $\v r$, $k_{\rm 0}=\omega\sqrt{\varepsilon_{\rm 0}\mu_{\rm 0}}$ is the free space wave number of the monochromatic field where $\varepsilon_{\rm 0}$ is the electric constant and $\alpha_{\rm m}$ is the magnetic polarizability. $\bar{\bar I}_{3}$ is an identity tensor. 
Although the free-space background is assumed in the proposed theory, a derivation assuming an inhomogeneous background can be achieved with the appropriate Green's function.

Combining Eq.~\eqref{eq:Ed} and Eq.~\eqref{eq:E_sca} gives
\begin{equation}
\label{eq:F-L_E}
    \v E_{i}=\v E^{\rm inc}_{i}+\omega^2\mu_{\rm 0}\sum_{j\neq i}^{N} \dy G_{ij}\alpha_{\rm e0}\v E_{j}.
\end{equation}
Here $\dy G_{ij}$ is shorthand for $\dy G_{ij}=\dy G(\v r_{i}, \v r_{j})$.
The particles are taken to be identical.
The electric field is then given by
\begin{equation}
\label{eq:F-L_inv}
    \ay E = \left(  \mx I - \mx G\alpha_{\rm e0}   \right)^{-1}\ay E^{\rm inc}
\end{equation}
Here $\ay E = \left(\v E_1, \v E_2 \ldots \v E_N \right)^T$ denotes a vector containing all $\v E_{i}$ on $N$ dipoles while $\mx G$ denotes the matrix containing all $\omega^2\mu_{\rm 0}\dy G_{ij}$ tensors. 
The identity matrix $\mx G$ is of the same dimension as $\mx G$.

The Foldy-Lax equation in \reqn{F-L_inv} provides a means for computing the field scattered by particles with electric polarizabilities, but not magnetic polarizabilities.
Atoms, molecules and particles can also have magnetic polarizability, and including such particles in a Foldy-Lax approach presents special challenges.
A novel approach to overcome these challenges and provide a solution to the associated generalization of the Foldy-Lax equation is presented in the following section.

\subsection{Generalized Foldy-Lax equations}
\label{subsec:G_FL}

Consider scattering from particles which respond also to magnetic fields. 
The magnetic current of the particle polarized by the magnetic field is given by
\begin{equation}{\label{eq:M}}
\v M(\v r, \v r_{i})  = i\omega\alpha_{\rm m0}\delta(\v r- \v r_{i})\v H(\v r_{i}).
\end{equation}
The electric and magnetic fields scattered from a particle are generated by both currents $\v J$ and $\v M$, given by~\cite{chew1995waves,sun2009novel,Cui2019ExceptionalCrystals}
\begin{widetext}

\begin{subequations}{\label{eq:EH_sca}}
\begin{align}
\label{eq:EH_sca_1}
\v E^{\rm sca}(\v r, \v r_{i}) & = \int i\omega\mu_{\rm 0} \dy G(\v r, \v r') \cdot \v J(\v r', \v r_{i}){\rm d}^3 r' + \int \dy G(\v r, \v r')\cdot \Big[ \nabla'\times \v M(\v r', \v r_{i}) \Big] {\rm d}^3 r', \, \v r \neq \v r_{\rm i},
\\
\label{eq:EH_sca_2}
\v H^{\rm sca}(\v r, \v r_{i}) & = \int -i\omega\varepsilon_{\rm 0} \dy G(\v r, \v r') \cdot \v M(\v r', \v r_{i}){\rm d}^3 r' + \int \dy G(\v r, \v r')\cdot \Big[ \nabla'\times  \v J(\v r', \v r_{i}) \Big] {\rm d}^3 r', \, \v r \neq \v r_{\rm i}.
\end{align}
\end{subequations}

\end{widetext}
Here $\nabla' \times$ denotes a curl operator acting on $\v r'$. The calculation of \reqn{EH_sca} is complicated by the point-particle assumption, which forces us to deal with a curl operator on a Dirac delta function. 
To deal with singularity introduced, the particle is diffused into a small volumetric distribution~\cite{DeVries1998PointWaves,born2013principles} with probability of finding a particle at point $\v r$:
\begin{equation}
\label{eq:P}
P(\v r, \v r_{i}) =\frac{1}{(4\pi D \Delta t)^{\frac{3}{2}}}{\rm exp}\left(-\frac{\lVert \v r - \v r_{i}\rVert^2}{4D\Delta t}\right).
\end{equation}
This equation describes a particle indexed by $i$ in Brownian motion around a location $\v r_i$ diffused for a time $\Delta t$ and diffusivity $D$ \cite{Einstein1905UberTeilchen,Einstein1956InvestigationsMovement}. 
Instead of a point, the electric and magnetic currents are taken to be given by the expected value of all the currents at randomized locations, given by
\begin{equation}{\label{eq:IM}}
\begin{split}
\overline{\v J (\v r, \v r_{i})} = & \int \v J(\v r, \v r') P(\v r', \v r_{i}) {\rm d}^3 r'
\\
= & -i\omega\alpha_{\rm e}P(\v r, \v r_{i})\v E(\v r),
\\
\overline{\v M (\v r, \v r_{i})} = & \int \v M(\v r, \v r') P(\v r', \v r_{i}) {\rm d}^3 r'
\\
= &  i\omega\alpha_{\rm m}P(\v r, \v r_{i})\v H(\v r),
\end{split}
\end{equation}
which are continuous and have a well-defined value of the curl.
Here the overline denotes averaging on all the possible configurations of the particle locations.
Here $\alpha_{\rm e}$ and $\alpha_{\rm m}$ are \textit{renormalized} polarizabilities taking the ``dipole fluctuation''~\cite{Barrera1988RenormalizedTheory} into account, see Appendix A for details.

The averaged currents in \reqn{IM} are justified as follows.
An actual measurement of an optical observable such as the Poynting vector can be calculated by its time average.
Assuming ergodicity, the time average is replaced by an ensemble average~\cite{mishchenko_multiple_2006}.
The latter is decomposed into a coherent flux $\overline{\v S_{\rm coh} (\v r)} = {\rm Re}[\overline{\v E (\v r)} \times \overline{\v H^* (\v r)}]/2$ and an incoherent part given by Eq.~(14) in Ref.~\cite{Mackowski2013DirectParticles}.
Here the coherent field $\overline{\v E (\v r)}$ is calculated by averaging over all configurations \cite{ishimaru_theory_1977} and $\overline{\v H (\v r)}$ is the magnetic analogue.
The coherent electromagnetic fields are generated by the configuration-averaged electric and magnetic currents, with the configuration and time-independent Green's function.
 
The electric and magnetic fields scattered by the currents in \reqn{IM} are likewise taken to be the expected value of the electric and magnetic fields averaged over the ensemble of particles in Brownian motion. 
The currents induced on a fixed particle, given in \reqn{I} and \reqn{M}, are recovered by \reqn{IM} in the $\sqrt{2D\Delta t}\to 0$ limit.
That is, the field scattered by a particle moving during an infinitesimal time is considered the same as the field scattered by a motionless particle.

The currents appearing in \reqn{EH_sca} are replaced with their averaged values given in \reqn{IM} and summed over all particles: 

\begin{equation}{\label{eq:IM_sum}}
\begin{split}
\v J(\v r) & = \sum^N_{i=1}\overline{\v J(\v r, \v r_{i})}= -i\omega\alpha_{\rm e}\sum\limits^N_{i=1} P(\v r, \v r_{i})\v E(\v r),
\\
\v M(\v r) & = \sum^N_{i=1}\overline{\v M(\v r, \v r_{i})}= i\omega\alpha_{\rm m}\sum\limits^N_{i=1} P(\v r, \v r_{i})\v H(\v r).
\end{split}
\end{equation}

\reqn{EH_sca} then becomes

\begin{widetext}

\begin{subequations}{\label{eq:EH}}
\begin{align}
\label{eq:EH_1}
\v E(\v r) & = \v E^{\rm inc}(\v r) + \dashint \left( i\omega\mu_{\rm 0} \dy G(\v r, \v r') \v J(\v r') + \dy G(\v r, \v r')\cdot \Big[ \nabla'\times \v M(\v r') \Big] \right) {\rm d}^3 r',
\\
\label{eq:EH_2}
\v H(\v r) & = \v H^{\rm inc}(\v r) + \dashint \left( -i\omega\varepsilon_{\rm 0} \dy G(\v r, \v r') \v M(\v r') + \dy G(\v r, \v r')\cdot \Big[ \nabla'\times \v J(\v r') \Big] \right) {\rm d}^3 r'.
\end{align}
\end{subequations}

\end{widetext}
The equation above is the generalized Foldy-Lax equation, for the first time to the best of our knowledge, to take the cross-terms into consideration, which couples the electric and magnetic responses of the particles. 
\reqn{EH} much like the Foldy-Lax \reqn{F-L_E} can be solved self-consistently so that the solution is exact and contains all orders of scattering.
Here  the integral $\dashint...{\rm d}^3 r'$ denotes the so-called principal volume integral~\cite{VanBladel1961SomeSpace}.
A detailed justification of solving the averaged field self-consistently and of the application of the principal volume can be found in Appendix A.

\subsection{Iterative solution of the generalized Foldy-Lax equation}
\label{half-space}

In order to self-consistently solve \reqn{EH}, we have to be able to calculate the curls of the currents $\nabla\times\v J(\v r)$ and $\nabla\times\v M(\v r)$. 
To simplify the calculation of the curl, we make use of the knowledge that system consists of a large number of particles with a plane wave incident from the exterior.
The field in the region of the particles will behave as if the particles form a continuous medium and so if the particles are confined to a half-space and the incident field is a plane wave, we anticipate that the field can be approximated as a plane wave.
Under these assumptions, we calculate the curls by discretization on a cubic grid.
At the center of the voxel $I$, the electric and magnetic fields $\v E$ and $\v H$ are denoted $\v E_{I}$ and $\v H_{I}$ and the Green's function $\dy G(\v r_{I}, \v r_{J})$ is denoted $\dy G_{IJ}$, so that \reqn{EH} may be written in discretized form (the detailed derivation can be found in Appendix B):
\begin{subequations}{\label{eq:EH_d3}}
\begin{align}
\label{eq:EH_d3_1}
\v E_{I}
= 
\v E^{\rm inc}_{I}
+
k_{\rm 0}^2\sum_{J\neq I}\dy G_{IJ}\left[ \frac{\rho\alpha_{\rm e}}{\varepsilon_{\rm 0}} +
\frac{\rho\alpha_{\rm m}n}{\eta\mu_{\rm 0}}  \right]\v E_{J}\Delta V,
\\
\label{eq:EH_d3_2}
\v H_{I}
= 
\v H^{\rm inc}_{I}
+
k_{\rm 0}^2\sum_{J\neq I}\dy G_{IJ}\left[ \frac{\rho\alpha_{\rm m}}{\mu_{\rm 0}} + \frac{\rho\alpha_{\rm e}\eta n}{\varepsilon_{\rm 0}} \right]\v H_{J}\Delta V,
\end{align}
\end{subequations}
where $\rho$ is the volume number density of the particles.
The refractive index $n$ is the ratio between the wavenumber of the plane wave that propagates in the medium and the wavenumber of the same wave but propagating in the free space, whereas $\eta$ is the ratio given by $x$ component of $\textbf{E}$ by the $y$ component of $\textbf{H}$, $\eta =  E  / (\eta_{\rm 0} H)$, where $\eta_0=\sqrt{\mu_{\rm 0}/\varepsilon_0}$.
The second terms in the square brackets in \reqn{EH_d3_1} and in \reqn{EH_d3_2} containing $\eta$ correspond to the electric field generated by the magnetic response of the particles and the magnetic field generated by the electric response of the particles, respectively.

The \reqn{EH_d3} are redundant, only one needs to be solved.
Thus the terms in the braces $[...]$ in \reqn{EH_d3_1} and in \reqn{EH_d3_2} are equivalent:
\begin{equation}
\label{eq:chi}
     \frac{\alpha_{\rm e}}{\varepsilon_{\rm 0}} +
\frac{\alpha_{\rm m}n}{\eta\mu_{\rm 0}}   =  \frac{\alpha_{\rm m}}{\mu_{\rm 0}} + \frac{\alpha_{\rm e}\eta n}{\varepsilon_{\rm 0}},
\end{equation}
which is solved to find
\begin{equation}
\label{eq:eta}
     \eta = \frac{\eta_{\rm 0}^2\alpha_{\rm e} - \alpha_{\rm m} \pm \sqrt{(\alpha_{\rm m}-\eta_{\rm 0}^2\alpha_{\rm e})^2 + 4\alpha_{\rm e}\alpha_{\rm m}n^2\eta_{\rm 0}^2}}{2\eta_{\rm 0}^2\alpha_{\rm e}n}.
\end{equation}
The $\pm$ should be chosen to be the sign of ${\rm Re}(\eta_{\rm 0}^2\alpha_e + \alpha_m)$.
The algorithm converges to $n=\eta=1$ if the wrong sign is chosen.

With $\eta$ given by \reqn{eta}, \reqn{EH_d3_2} may be seen to be redundant.
\reqn{EH_d3_1} requires the refractive index $n$, which may be extracted from the calculated field distribution by
\begin{equation}
\label{eq:n}
    n \approx \frac{{\rm arg} (E_{I+1})- {\rm arg}( E_{I})}{k_{\rm 0} \Delta z}.
\end{equation}

\noindent
Here the indices $I+1$ and $I$ denote neighbouring voxels arranged along the wave propagation direction and $\Delta z$ is the distance between their centers.

\begin{figure}
\includegraphics{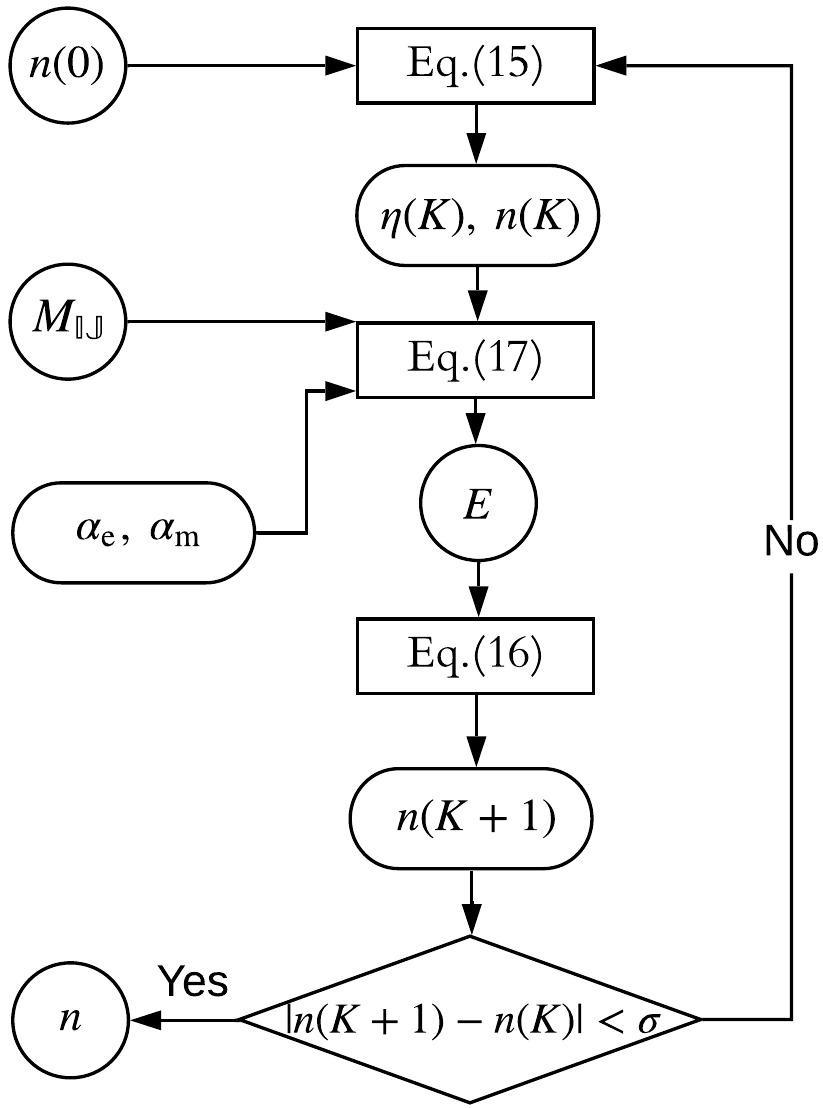}
\caption{\label{fig:main}
The diagram of the iterative solution of the generalized Foldy-Lax equation. 
With an initial guess $n(0)$, the electric field is calculated by the Foldy-Lax equation in lv3, \reqn{E_lv2}. 
A new value of the refractive index $n(K)$ in the $K$-th iteration is calculated by \reqn{n}, which is compared with the value from the last iteration until their difference is smaller than the threshold.}
\end{figure}

We propose an iterative algorithm for the calculation of $E$ and $n$. 
The electric field is calculated by \reqn{EH_d3_1} with an initial guess of refractive index $n(0)$. 
Then a new refractive index $n(1)$ is calculated from the electric field by \reqn{n}, and $n(1)$ is subsequently used in \reqn{EH_d3_1} to again calculate the electric field. 
This process is repeated until we reach the convergence criterion in the $K$-th iteration, $|n(K) - n(K-1)|/|n(K-1)|<\sigma$.
The user-defined value of $\sigma$ varies depending on specific applications, which is chosen to be $0.1\%$ in this paper.
This is discussed in \rfig{main} and below.

\subsection{The hierarchical clustering technique}
\label{subsec:clustering}

\begin{figure}
\includegraphics{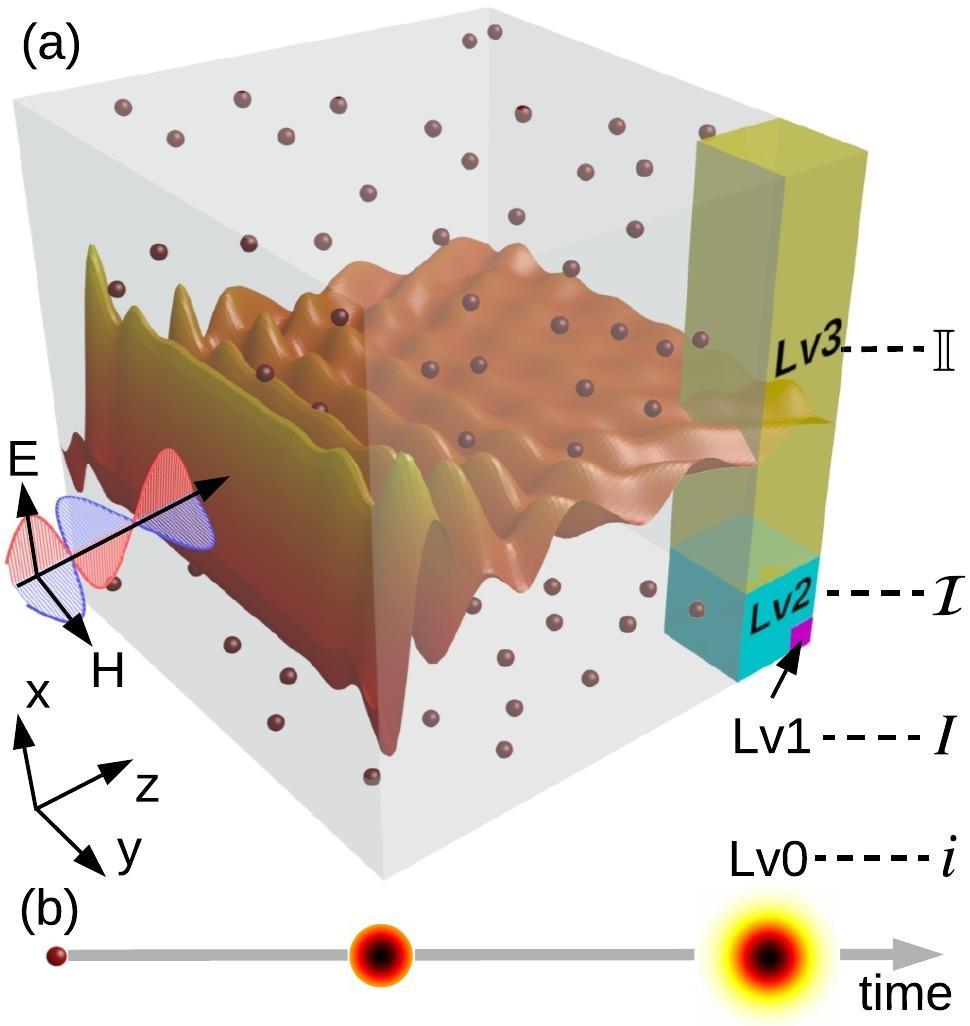}
\caption{\label{fig:Simu}
The simulation model of the diffused-particle method. 
(a) A total number of $N$ particles labeled $i$ with an electric and a magnetic response are randomly distributed within the simulation box.
The particles are clustered into lv1 voxels labeled $I$ and lv2 voxels labeled $\mathcal{I}$, which are further clustered into lv3 columns labeled $\mathbb{I}$, see table~\ref{tab:lv} for details.
The effective refractive index of the composed medium is extracted from the simulated electric field demonstrated in the middle plane of the simulation region.
(b) Each particle being a lv0 scattering unit is diffused. 
}
\end{figure}

\begin{table}[]
\begin{tabular}{|c|c|c|c|}
\hline
level & structure & index     & size parameters   \\ \hline
0           & diffused particle & $i$  & $\sqrt{2D\Delta t}$  \\ \hline
1           & small voxel & $I$  & $\Delta L_{\rm 1}\times\Delta L_{\rm 1}\times\Delta L_{\rm 1}$  \\ \hline
2  & larger voxel & $\mathcal{I}$ & $\Delta L_{\rm 2}\times\Delta L_{\rm 2}\times\Delta L_{\rm 2}$   \\ \hline
3  & column & $\mathbb{I}$ & $\Delta L_{\rm 2}\times\Delta L_{\rm 2}\times L$    \\ \hline
\end{tabular}
\caption{\label{tab:lv}
The structures, indices and parameters in different levels.
}
\end{table}

\begin{figure}
\includegraphics{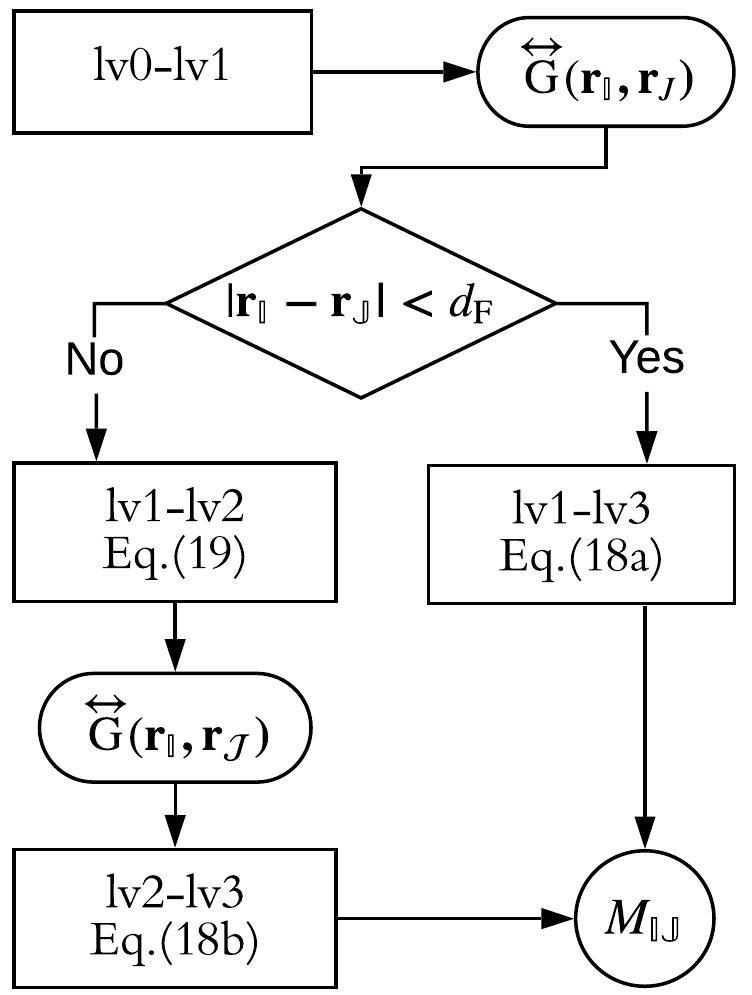}
\caption{\label{fig:preprocessing}
The diagram of calculating $M_{\mathbb{I}\mathbb{J}}$ describing the interaction between columns.}
\end{figure}

The analytic description of an infinite number of particles in a semi-infinite half-space given above provides a means for calculating the effective refractive index, however numerics must be carried out over a finite number of particles, and so we confine our attention to a finite-sized box as described below.
The simulation region is taken to be a cube shown in \rfig{Simu}(a). 
The $N$ particles are diffused into volumetric currents as described in Sec.~\ref{subsec:G_FL}.
The diffusion of a single particle is illustrated in \rfig{Simu}(b). 
Although solving the interaction of $N$ particles is a challenge, the hierarchical clustering method, with the concept inspired by works in other research fields~\cite{Song1997MLFMAObjects,Hackbusch1999A-Matrices,shih_coarse_2006,shih_assembly_2007,arkhipov_stability_2006,arkhipov_four-scale_2008,corpet_multiple_1988,johnson_hierarchical_1967}, reduces the computational complexity from $O(N^3)$ to $O(N_{\rm top}^3)$, where $N_{\rm top}$ is the number of top level structures. 
With a diagram given in \rfig{preprocessing}, the calculations are performed in four levels: at Level 0 (lv0), fields scattered from individual particles are calculated; at Level 1 (lv1) particles are clustered into cubes with homogeneous permittivity and permeability such that the field scattered from that cube matches the field scattered from lv0 particles within a cube in a minimum $\ell_2$-norm error sense; at Level 2 (lv2), the cubes are clustered into larger cubes treated as homogeneous so the field scattered from a lv2 voxel is equivalent to the field scattered from all the lv1 voxels within the lv2 voxel; at Level 3 (lv3), the lv2 voxels are clustered into columns, which again are treated as homogeneous, and fields scattered from these columns are calculated.
The side length of the lv2 voxel is equivalent to the cross-section side length of the lv3 column.
The fields scattered by individual particles, that is the calculation at lv0, are described by \reqn{EH}.
The lv1 and lv2 fields, those are fields scattered by the cubes, are given by \reqn{EH_d3_1}.
The lv3 fields, those are fields scattered by the columns, are given by
\begin{equation}
\label{eq:E_lv2}
E_{\mathbb{I}}
= 
E^{\rm inc}_{\mathbb{I}}
+
k_{\rm 0}^2\sum^{N_{\rm 3}}_{\mathbb{J}=1}M_{\mathbb{I}\mathbb{J}}\left[ \frac{\rho\alpha_{\rm e}}{\varepsilon_{\rm 0}} +
\frac{\rho\alpha_{\rm m}n}{\eta\mu_{\rm 0}}  \right] E_{\mathbb{J}}\Delta S_{\rm 2}.
\end{equation}
Here $\Delta S_{\rm 2} = (\Delta L_{\rm 2})^2$ denotes the area of the lv3 column on the $yz$ plane, where $\Delta L_{\rm 2}$ is the side length of a lv2 voxel, and $M_{\mathbb{I}\mathbb{J}}$ represents the re-summed discretized Green's function for scattering from column $\mathbb{J}$ to column $\mathbb{I}$ at lv3 of a clustering procedure.
We assume that near the center of the simulation volume the field propagating in the medium is, to a good approximation, plane-wave-like, see Fig.~\ref{fig:Simu}(a).
With this assumption and the medium being isotropic, only the $xx$ component of $M_{\mathbb{I}, \mathbb{J}}$ needs to be calculated for the incident field polarized along $\hat{x}$ direction, as then the field propagating in the medium must also be polarized along $\hat{x}$ direction.
Thus, we see the benefit of the clustering approach: a problem set in 3 spatial dimensions in \reqn{EH_d3_1} is reduced to a problem set in 2 spatial dimensions in \reqn{E_lv2}.

The calculation of $M_{\mathbb{I}\mathbb{J}}$ is carried out in 2 different ways, with the diagram given in \rfig{preprocessing}.
The choice of which technique to use depends on the distance $|\v R| = |\v r_\mathbb{I} - \v r_\mathbb{J}|$. 
The limit between the near field, $|\v R| \leq d_{\rm F}$, and far field, $|\v R|>d_{\rm F}$, is denoted as $d_{\rm F}$.
The value of $d_{\rm F}$ can be pre-calculated as described in Appendix C.
In the \textit{near field} region, where the fine mesh is required, the calculation of $M_{\mathbb{I}\mathbb{J}}$ takes the lv0-lv1-lv3 hierarchical clustering procedure, while in the \textit{far field} region, where the coarse approach is applicable, lv0-lv1-lv2-lv3 clustering is used:
\begin{numcases}{M_{\mathbb{I}\mathbb{J}} = }
    \sum_{J} \left[ \dy G(\v r_{\mathbb{I}}, \v r_{J}) \right]_{xx} \frac{\Delta V_{\rm 1}}{\Delta S_{\rm 2}},&$|\v R|\leq d_{\rm F}$ \label{eq:M_op}
    \\
    \sum_\mathcal{J} \left[ \dy G(\v r_{\mathbb{I}}, \v r_\mathcal{J}) \right]_{xx} \Delta L_{\rm 2},&$|\v R|>d_{\rm F}$ \label{eq:M_op1}
\end{numcases}
where
\begin{equation}
\label{eq:G_lv2}
 \left[ \dy G(\v r, \v r_\mathcal{J}) \right]_{xx} \Delta V_{\rm 2} \approx \sum_{J} \left[ \dy G(\v r, \v r_{J}) \right]_{xx}\Delta V_{\rm 1}.
\end{equation}
Here $J$, the index of a lv1 voxel, ranges over all possible values within the lv2 voxels labelled by the index $\mathcal{J}$, which runs over all possible values within the lv3 column labelled by $\mathbb{J}$. 
The volume of a lv1 voxel is given by $\Delta V_{\rm 1} = (\Delta L_{\rm 1})^3$, where $\Delta L_{\rm 1}$ is the side length of a lv1 voxel.
Introducing the lv2 voxels reduces computational complexity by reducing the number of scatterers interacting with each other.

\section{Results and Discussion}
\label{numerical}

\subsection{The simulated field and refractive index}

In order to calculate the effective refractive index from \reqn{n}, we need to simulate the scattered field in a cubic volume large enough so that in a region near the center of a cube the field behaves as if it propagates in a semi-infinite medium.
On the other hand, the numerical complexity scales as $L^6$ where $L$ is a cube length, given the same discretization of the simulation region, i.e., the size of the lv1 voxel.
The size of the cubic simulation volume in \rfig{Simu} is selected to balance the accuracy and complexity of the calculation. 
By repeated numerical experimentation, we found that a cube with a size of $L=4.2\lambda$ on a side allows us to calculate the effective refractive index while running a reasonable time ($<1$ hours on a $2.3$ GHz Intel Core i5 CPU), where $\lambda$ is the wavelength of the field in free-space.
The total number of particles is chosen to be $N = 4\times 10^{10}$ which corresponds to the atomic density of silicon for $\lambda = 221$~nm at a temperature of $300$~K and under a pressure of $1$~atm.

The size of lv1 and lv2 voxels is determined as follows.
The scattered field arises from the interaction of particles and the propagating field.
The rapid variation of the susceptibilities can be ignored while the size of the voxels should be of the scale of the variations of the propagating field.
So the side length of a lv1 voxel should be much smaller than the wavelength of the field propagating in the medium (i.e., less than $0.01\lambda/n$).
Here we specifically assign the side length of a lv1 voxel as $0.0015\lambda$.
The side length of the lv2 voxel is chosen to be $0.03\lambda$, justified by \rfig{lv012er} in Appendix D.
Indeed, increasing the size of the lv2 voxel to include more lv1 voxel reduces complexity.
However, the error of the lv1-lv2 clustering procedure, plotted in \rfig{lv012er}, also increases, because of the non-trivial higher order of the multipole components of the voxel~\cite{darve_fast_2000,engheta_fast_1992,coifman_fast_1993}.
Thus the size of the lv2 voxel is limited for the purpose of the accuracy of the calculation.
Given this size and the concentration chosen, each lv2 voxel comprises $14600$ particles.
The $N_{\rm 2}=2.74\times10^6$ lv2 voxels at locations specified by coordinates $y$ and $z$ are clustered into $N_{\rm 3}=1.96\times 10^4$ columns, each consisting of $140$ lv2 voxels.
The interaction matrix elements between columns, $M_{\mathbb{I}\mathbb{J}}$ is calculated as shown in the flow chart, \rfig{preprocessing}.
Much like a Green's function in free space, $M_{\mathbb{I}\mathbb{J}}$, depends only on geometry and wavelength in free space, not on the refractive index or impedance.
With $M_{\mathbb{I}\mathbb{J}}$ pre-calculated for our chosen hierarchical clustering process \footnote{The operator $M$ is a Toeplitz matrix if hexahedron mesh is used in the simulation. Taking advantage of the fact that the elements are repeated saves both RAM and CPU time.}, the effective refractive index and the electric field distribution in the simulation region are calculated using the iteration scheme given in \rfig{main}.

With a goal in mind to find a material with an effective refractive index with a real part of 2, we take the particles polarizabilities to be $\alpha_{\rm e}= (1.41\times10^{-9}+1.31\times10^{-10}i)\lambda^3\varepsilon_{\rm 0}$ and $\alpha_{\rm m}= (5.62\times10^{-10}+1.78\times10^{-11}i)\lambda^3\mu_{\rm 0}$, starting from these values of polarizabilities, we can calculate the effective refractive index and the electric field distribution throughout the simulation region.
To start the iteration process, the initial guess of the refractive index is taken to be the same as the free-space, $n(0)=1$.
The convergence criterion, $|n(K) - n(K-1)|/|n(K-1)|< 0.1 \%$, is met in the $9$th iteration, with the resultant refractive index $2.0 + 0.2i$.
The simulated electric field in the last iteration is shown in the middle plane in \rfig{Simu}.
It may be observed that this electric field is a superposition of the ideal planewave to which we fit to calculate the index of refraction plus the deviations from that ideal field generated by the boundaries.

\begin{figure}
\includegraphics{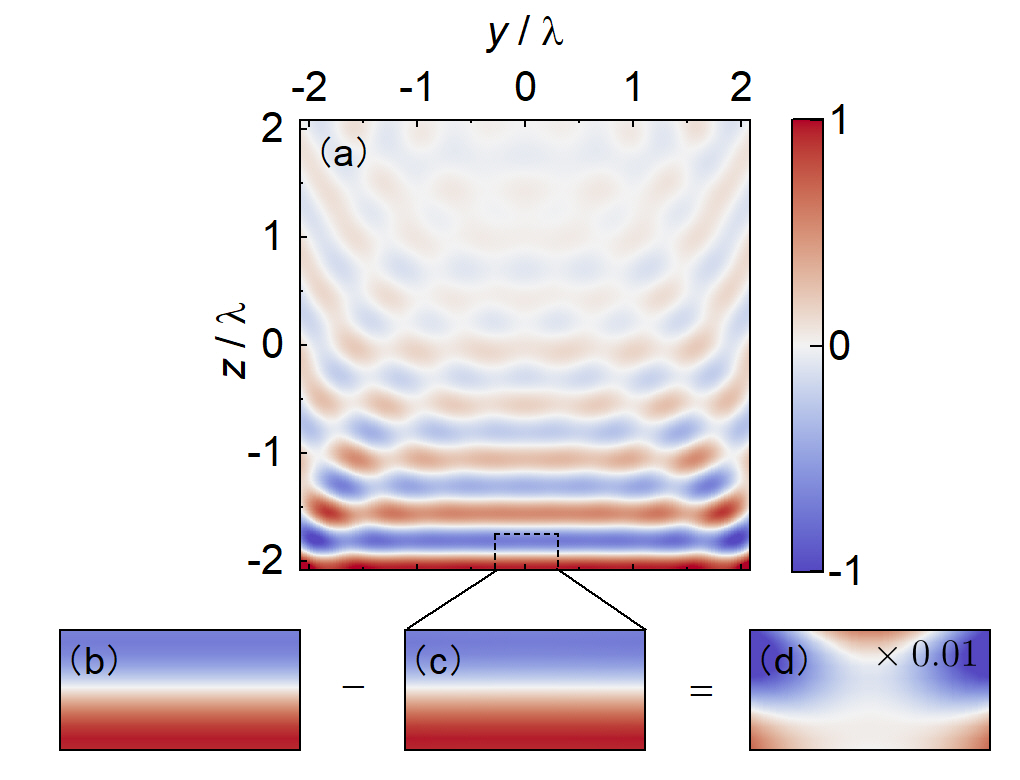}
\caption{\label{fig:sample}
(a) The simulated scattered field as described in the text, (b) the ideal plane wave used to fit (c) the computed field in the $0.6\lambda \times 0.3\lambda$ region marked with black dotted-line rectangle in (a). (d) The difference between the simulated total field and the plane wave with the color bar scaled by $0.01$.}
\end{figure}

Though the electric field and the refractive index have been acquired, the validity of \reqn{n} used to calculate the refractive index must be checked \textit{post hoc}.
\rfig{sample} shows the deviation from the ideal planewave used to fit the field and extract the index.
We see that in the region used to calculate the refractive index the planewave dominates in the sense that $\langle |{\rm arg}( E^{\rm plane})- {\rm arg}( E^{\rm simu})|/{\rm arg}( E^{\rm plane}) \rangle <0.1\%$, where the $\langle ... \rangle$ denotes averaging throughout the rectangle.
The error of calculating the refractive index with \reqn{n} is thus limited to less than $0.2 \%$.

\subsection{Comparing the numerical and analytical results}

\begin{figure}
\centering
\includegraphics{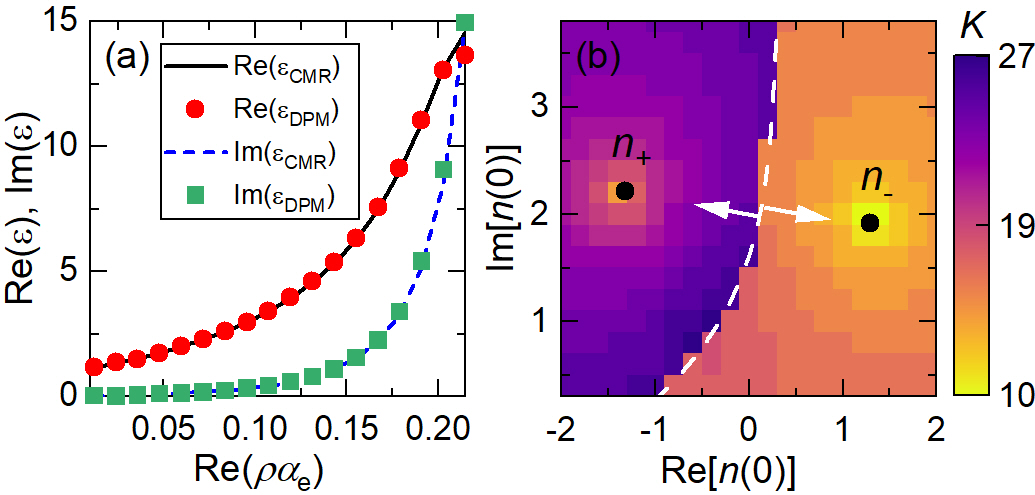}
\caption{\label{fig:CMconv}
(a) The comparison of the theoretical $\varepsilon_{\rm CMR}$ calculated from \reqn{CM} and simulated permittivity $\varepsilon_{\rm DPM}$ of an effective medium composed of particles with pure electric response. 
(b) The number of iterations required for convergence of the simulated refractive index to either branch, $n_+$, or, $n_-$, predicted by the generalized Clausius-Mossotti relation~\cite{Lang_Ilia_Scott_1}, depending on which side of the curve the initial value $n(0)$ falls on.}
\end{figure}

Having established self-consistency, let us here compare the results of our diffused particle method with well-known results from the Clausius-Mossotti relation~\cite{clausius1879mechanical,mossotti1850discussione,lorenz1869experimentale,lorentz1909theory,Thackray2014,Markel2005}, that is we hold the magnetic polarizability to be zero so that the usual Clausius-Mossotti relation applies.
We apply procedures as described above.
The particles are chosen to be lossy, with polarizability ${\rm Im}(\alpha_{\rm m}) = 0.1 {\rm Re}(\eta_{\rm 0}^2\alpha_{\rm e})$.
With $\alpha_{\rm m}=0$, the term proportional to the curl of magnetic currents in \reqn{EH} vanishes.
As may be seen in \rfig{CMconv}(a), the simulated permittivity agrees well with the theoretical result given by the Clausius-Mossotti relation,
\begin{equation}
\label{eq:CM}
    \frac{\rho\alpha_{\rm e}}{3\varepsilon_{\rm 0}} = \frac{\varepsilon -1}{\varepsilon + 2}.
\end{equation}

Then the diffused particle method is used to calculate the effective refractive index of the medium comprising particles responding to both electric and magnetic field, with $\alpha_{\rm e} = (5.55\times 10^{-9}+5.55\times 10^{-10}i)\lambda^3 \varepsilon_{\rm 0}$ and $\alpha_{\rm m} = (5.55\times 10^{-10}+5.55\times 10^{-11}i)\lambda^3 \mu_{\rm 0}$.
In this case, the generalized Clausius-Mossotti relation~\cite{Lang_Ilia_Scott_1} gives 2 different values of the effective refractive index, $n_{\rm +}$ and $n_{\rm -}$.
The simulated refractive index in \rfig{CMconv}(b) converges to either predicted value depending on the initial guess $n(0)$.
The relative difference between the theoretical refractive index and the simulated value is less than $0.2 \%$.

\section{Conclusion}

Nature rarely produces atoms or molecules with a magnetic permeability, but not never.
When electric and magnetic polarizabilities co-exist on the same particle, they must naturally interact. 
We have for the first time presented a generalized Foldy-Lax relation for simultaneously electric and magnetic polarizabilities.

We have presented numerical methods that are used to calculate the effective permittivity and permeability of a medium composed of more than $10^{10}$ particles with both electric and magnetic responses simultaneously.
Corresponding MATLAB routines, which include the theoretical treatment reported in this paper, are presented in Ref.~\cite{noauthor_httpsgithubcomlwang111diffused-particle-method_nodate}.
Using this new method, we have validated analytical results generalizing the Clausius-Mossotti relation to such materials.
At the heart of this numerical method, a generalized Foldy-Lax equation is derived to calculate the field distribution among the particles. 
The numerical solution is achieved by applying hierarchical clustering techniques. 
Macroscopic optical properties of an effective continuous medium equivalent to the collection of particles are computed from the numerical results for the field.
These macroscopic results agree well with the analytical results provided by effective medium theories.

The method used to calculate the effective medium parameters from the numerical results depends on an assumption that the field behaves locally as a planewave, an assumption that works well for a lossy medium.
Of course, one could instead compute the field expected for a continuous medium of the same size and shape as the simulation volume for our collection of particles and then fit the macroscopic properties of that medium to the numerical results.
This approach is left to future work and will require another iterative algorithm to find the local minimum of the cost function~\cite{Chew1990ReconstructionMethod}.

The particle clustering techniques used here to homogenize the medium may fail in certain cases.
For example, for arbitrarily shaped materials or low loss or gain medium, different clustering techniques are suggested~\cite{Song1997MLFMAObjects,Hackbusch1999A-Matrices}.
Although materials with randomly distributed particles are chosen for the calculation in this paper, crystal structures with naturally periodic discretization can also be calculated by the proposed method. 
The process of clustering the particles to a voxel is unnecessary in a crystal structure, while the rest of the steps are the same as the random particle distribution case.

A number of intriguing avenues of the investigation remain.
We have only considered particles with positive real polarizabilities, but of course particle polarizability with a negative real part is possible~\cite{Lewin1947TheParticles,Khizhniak1957I,Holloway2003AMatrix,Zhao2009MieMetamaterials}, and might open a broader parameter space with more opportunities to find materials with exotic electromagnetic responses.
Nonlinear and multipolar polarizabilities of the particles are omitted in the derivation of the generalized Foldy-Lax equation, which is the subject of further research.

\begin{widetext}

\appendix
\renewcommand{\theequation}{A\arabic{equation}}
  \setcounter{equation}{0}
\section*{Appendix A: the validation of self-consistently solving the configuration-averaged fields}

\begin{figure}
\includegraphics[width=2in]{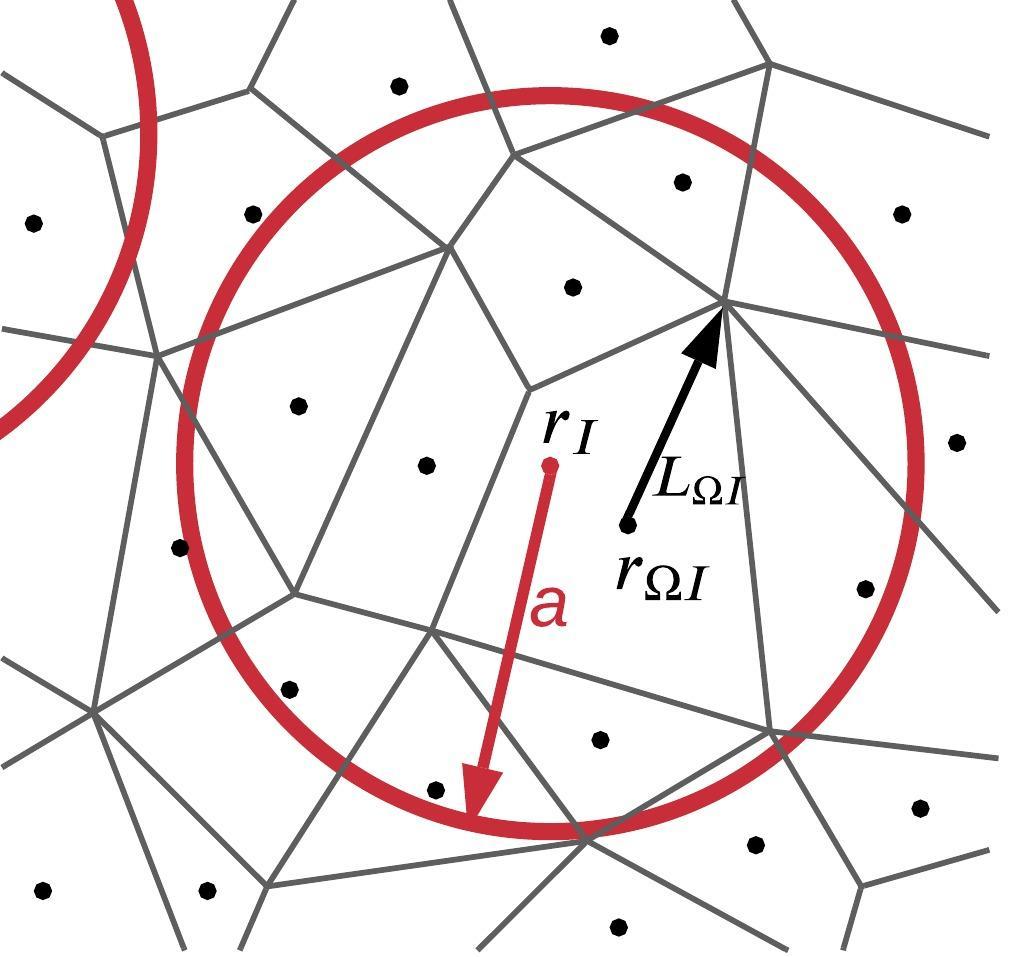}
\caption{\label{fig:avg}
An illustration of a cell $\Omega_I$ centered at $\v r_{\Omega I}$ and a hard sphere of a dipole. The hard sphere is centered at the dipole located at $\v r_I$ inside $\Omega_I$. $L_{\Omega I}$ denotes the longest distance between $\v r_{\Omega I}$ and any point in $\Omega_I$, $a$ is the radius of the hard sphere of a dipole. The condition $\lambda \gg a \gg L_{\Omega I}$ is satisfied in all cases.}
\end{figure}

In this Appendix, we justify the validity of \reqn{IM_sum} and \reqn{EH} as the configuration-averaged currents.
To calculate an averaged current distribution in a medium consisting of randomly distributed particles, it is common to conduct a Monte Carlo simulation where the currents are solved in each configuration of particles locations and then averaged over all configurations.
In this paper, however, a configuration-averaging is performed over the currents in \reqn{IM_sum} and \reqn{EH}, and the averaged currents are then solved self-consistently.
This approach eliminates the calculation of a curl of the singularity introduced by the point-particle assumption in \reqn{I} and \reqn{M}.
It also reduces the computational complexity because the averaged currents are only calculated once in the whole process.
For simplicity, we ignore the magnetic response of the particles in this Appendix, i.e. $\alpha_{\rm m}=0$.
The general case of particles responding to both electric and magnetic fields follows a similar pattern.

We discretize the space into $N_{\rm c}$ cells denoted by $\Omega_I$ indexed by $I\in[1,N_{\rm c}]$.
We calculate the current at the center of each cell (the mass center of the cell with a uniform density), $\v r_{\Omega I}$, see \rfig{avg}.
The cells and the centers $\v r_{\Omega I}$ remain the same in different configurations.
The longest distance between $\v r_{\Omega I}$ and any point in $\Omega_I$ is denoted by $L_{\Omega I}$, and $L_{\rm max} \equiv \max_I(L_{\Omega I})$ is defined as the maximum $\v r_{\Omega I}$ among all cells. 
The total electric current in cell $\v \Omega_I$ obeys the Foldy-Lax equation 

\begin{equation}
\label{eq:J_particle}
    \v J_I = \v J^{\rm inc}_I+\omega^2\mu_{\rm 0}\alpha_{e0} N_I\sum_{J=1,J\neq I}^{N_{\rm c}}\dy G(\v r_I, \v r_J)\cdot  \v J_J,
\end{equation}
\noindent
where $N_I$ denotes the number of dipoles in $\Omega_I$, which takes the value $0$ in most cells without a particle inside and the value $1$ with a particle located at $\v r_I\in \Omega_I$; $\v J^{\rm inc}_I = N_I\alpha_{\rm e} \v E^{\rm inc}_I$ is the total current in cell $\Omega_I$ induced by the incident field.
Following previous simulation methods~\cite{mishchenko_multiple_2006,Mackowski2013DirectParticles}, we assume the particles behave as hard spheres.
The self-interaction of a particle with the field scattered from itself is already taken into account in the polarizability of the particle, \reqn{I}.
We use such a discretization which guarantees $L_{\rm max} \ll a$, that is the discretization is much smaller than the size of hard spheres representing the particles, so, at most, one dipole is found in each $\Omega_J$.
Thus we can index the locations of particles $\v r_I$ and $\v r_J$ in \reqn{J_particle} by the index of the cells containing the dipoles $I$ and $J$.

To prove that the configuration-averaged currents can be solved self-consistently, we need to show that the current at $\v r_I$ (the location of a particle in cell $\Omega_I$) can be approximated by replacing the dyadic Green's function $\dy G(\v r_I, \v r_J)$ from dipole to dipole by the dyadic Green's function $\dy G(\v r_{\Omega I}, \v r_{\Omega J})$ from cell to cell

\begin{equation}
\label{eq:J_cell}
    \v J_I \approx \v J^{\rm inc}_I+\omega^2\mu_{\rm 0}\alpha_{e0} N_I\sum_{J=1,J\neq I}^{N_{\rm c}}\dy G(\v r_{\Omega I}, \v r_{\Omega J})\cdot  \v J_J,
\end{equation}
with an error under proper control.

The matrix-vector forms of \reqn{J_particle} and \reqn{J_cell} are:

\begin{equation}
\label{eq:J_particle_matrix}
\ay J = \ay J^{\rm inc} +(\mx G+\Delta \mx G)\ay J,
\end{equation}

\begin{equation}
\label{eq:J_cell_matrix}
\ay J \approx \ay J^{\rm inc}+\mx G \ay J.
\end{equation}

\noindent
Here $\ay J$ is a vector of vectors denoting the total currents in all cells, with the $I$th vector as $\v J_I$;
$\ay J^{\rm inc}$ denotes the current induced by the incident fields in all cells, with the $I$th vector as $\v J^{\rm inc}_J$;
$\mx G$ denotes an operator of tensors with the $IJ$th tensor as $\omega^2\mu_{\rm 0}\alpha_{e0} N_J \dy G(\v r_{\Omega I}, \v r_{\Omega J})$;
$\Delta \mx G$ denotes an operator of tensors with the $IJ$th tensor given by

\begin{equation}
\label{eq:D_G}
    \Delta \mx G_{IJ} \equiv \omega^2\mu_{\rm 0}\alpha_{e0} N_J[\dy G(\v r_I, \v r_J) - \dy G(\v r_{\Omega I}, \v r_{\Omega J})].
\end{equation}

\noindent
The error between the exact $\ay J$ calculated by \reqn{J_particle_matrix} and the approximated $\ay J$ by \reqn{J_cell_matrix} is defined as

\begin{equation}
\label{eq:J_er}
\Delta \ay J^{\rm er} \equiv (\mx I - \mx G-\Delta \mx G)^{-1}\ay J^{\rm inc} - (\mx I - \mx G)^{-1}\ay J^{\rm inc}= \mx F \ay J^{\rm inc},
\end{equation}

\begin{equation}
\label{eq:F}
    \mx F = \Delta \mx G + \mx G\Delta \mx G +\Delta \mx G \mx G + \mx G \mx G\Delta \mx G + \mx G\Delta \mx G \mx G + \Delta \mx G \mx G \mx G+...+\mathcal{O}[(\Delta \mx G)^2].
\end{equation}

\noindent
The norm of the error current is bounded by

\begin{equation}
\label{eq:J_er_norm}
\lVert \Delta \ay J^{\rm er} \rVert \leq \lVert \mx F \rVert \lVert \ay J^{\rm inc} \rVert.
\end{equation}
Here $\lVert ... \rVert$ denotes the 2-norm of the vector when acting on a vector and denotes the 2-norm of a matrix induced by the 2-norm of a vector when acting on a matrix.
The 2-norm of a matrix is defined as

\begin{equation}
\label{eq:norm}
\lVert \mx A \rVert = \sup_{\ay x\neq 0}\frac{\lVert \mx A \ay x \rVert}{\lVert \ay x \rVert}.
\end{equation}

The norm of the vector of the currents induced by the incident field is bounded and remains the same when the sizes of cells decrease. The norm of operator $\mx F$ is bounded by

\begin{align}
\label{eq:F_norm}
    \lVert \mx F\rVert & \leq  \lVert \Delta \mx G\rVert + \lVert \mx G\Delta \mx G\rVert +\lVert \Delta \mx G \mx G\rVert + \lVert \mx G\mx G\Delta \mx G\rVert + \lVert \mx G\Delta \mx G \mx G\rVert + \lVert \Delta \mx G\mx G\mx G\rVert+...+\lVert \mathcal{O}[(\Delta \mx G)^2]\rVert\\
    & \leq \lVert \Delta \mx G\rVert + 2\lVert \mx G\rVert\lVert \Delta \mx G\rVert  + 3\lVert \mx G\rVert\lVert \mx G\rVert\lVert \Delta \mx G\rVert +...+\lVert  \mathcal{O}[(\Delta \mx G)^2]\rVert \\
    & = \frac{\lVert \Delta \mx G\rVert}{(1-\lVert  \mx G\rVert)^2}+\lVert  \mathcal{O}[(\Delta \mx G)^2]\rVert.
\end{align}
We require that $\lVert \mx G\rVert \neq 1$. The norm of the operator $\lVert \Delta \mx G \rVert$ is bounded by

\begin{equation}
\label{eq:Delta_Gs_limit}
\begin{split}
\lVert \Delta \mx G \rVert & \leq \omega^2\mu_{\rm 0}\alpha_{\rm e} \sum_{I,J}\lVert \Delta \mx G_{IJ}\rVert \\
& = \omega^2\mu_{\rm 0}\alpha_{\rm e} \sum_{I,J} \lVert \dy G(\v r_I, \v r_J) - \dy G(\v r_{\Omega I}, \v r_{\Omega J})\rVert \\
& \leq \omega^2\mu_{\rm 0}\alpha_{\rm e}N_{\rm d}(N_{\rm d}-1) \max_{I,J}(\lVert \dy G(\v r_I, \v r_J) - \dy G(\v r_{\Omega I}, \v r_{\Omega J})\rVert),
\end{split}
\end{equation}

\noindent
which is the multiplication of the number of non-zero tensors, $N_{\rm d}(N_{\rm d}-1)$, and the maximum norm of its tensor elements $\lVert \dy G(\v r_I, \v r_J) - \dy G(\v r_{\Omega I}, \v r_{\Omega J})\rVert$.

The norm of the difference between two Green's tensors $\lVert \dy G(\v r_I, \v r_J) - \dy G(\v r_{\Omega I}, \v r_{\Omega J})\rVert$ is bounded by the following process.
The value of the dyadic Green's function in free-space from a source point $\v r_{\Omega J}$ to an observation point $\v r_{\Omega I}$ is only determined by the displacement of the two points $\v R \equiv \v r_{\Omega I} - \v r_{\Omega J}$:

\begin{equation}
\label{eq:G_IJ}
\dy G(\v r_{\Omega I}, \v r_{\Omega J}) =\dy G(\v R) = \left[ \left(\frac{1}{k_{\rm 0}R} + \frac{i}{(k_{\rm 0}R)^2} - \frac{1}{(k_{\rm 0}R)^3} \right)\bar{\bar I}_{3} + \left(-\frac{1}{k_{\rm 0}R} - \frac{3i}{(k_{\rm 0}R)^2} + \frac{3}{(k_{\rm 0}R)^3} \right) \hat{\v R} \otimes \hat{\v R} \right]\frac{k_{\rm 0}{\rm exp}(i k_{\rm 0} R)}{4\pi}.
\end{equation}

\noindent
Here $\hat{\v R} \equiv \v R / R$ is a unit vector, and $\otimes$ denotes a tensor product.
We introduce $\Delta \v R = (\v r_I-\v r_J)-\v R$, then

\begin{equation}
\label{eq:Delta_G}
\begin{split}
& \dy G(\v r_I, \v r_J) - \dy G(\v r_{\Omega I}, \v r_{\Omega J}) \\
= & \left\{\left[ \frac{i}{k_0 R}-\frac{2}{(k_0 R)^2}-\frac{3i}{(k_0 R)^3}+\frac{3}{(k_0 R)^4} \right]\mx I_3+\left[ -\frac{i}{k_0 R}+\frac{4}{(k_0 R)^2}+\frac{9i}{(k_0 R)^3}-\frac{9}{(k_0 R)^4} \right]\hat{\v R} \otimes \hat{\v R}\right\}\frac{k_0{\rm exp}(i k_0 R)(k_0 \Delta R)}{4\pi} \\
& + \left(-\frac{1}{k_{\rm 0}R} - \frac{3i}{(k_{\rm 0}R)^2} + \frac{3}{(k_{\rm 0}R)^3} \right)\frac{k_0{\rm exp}(i k_0 R)}{4\pi}\left( \frac{\Delta \v R}{R} \otimes \hat{\v R} + \hat{\v R} \otimes \frac{\Delta \v R}{R} \right)+k_0\mathcal{O}[(k_0\Delta R)^{2}]
\end{split}
\end{equation}

\noindent
Since the length of the displacement $\Delta R\leq 2L_{\rm max}$, the norm of the difference between the two Green's tensor in the equation above is bounded by

\begin{equation}
\label{eq:Delta_G_lim_1}
\lVert \dy G(\v r_I, \v r_J) - \dy G(\v r_{\Omega I}, \v r_{\Omega J}) \rVert 
\leq \left[ \frac{1}{k_0 R}+\frac{4}{(k_0 R)^2}+\frac{9}{(k_0 R)^3}+\frac{9}{(k_0 R)^4} \right]\frac{k_0(k_0 L_{\rm max})}{\pi}+k_0\mathcal{O}[(k_0L_{\rm max})^{2}].
\end{equation}

\noindent
The length of vector $\v R$ is longer than the minimal distance between two dipoles, $R \geq 2a$.
Under this condition, we have

\begin{equation}
\label{eq:Delta_G_lim_2}
\lVert \dy G(\v r_I, \v r_J) - \dy G(\v r_{\Omega I}, \v r_{\Omega J}) \rVert 
\leq \left[ \frac{1}{2k_0 a}+\frac{4}{(2k_0 a)^2}+\frac{9}{(2k_0 a)^3}+\frac{9}{(2k_0 a)^4} \right]\frac{k_0(k_0 L_{\rm max})}{\pi}+k_0\mathcal{O}[(k_0L_{\rm max})^{2}].
\end{equation}

\noindent
for any $I$ and $J$. Combining \reqn{Delta_Gs_limit} and \reqn{Delta_G_lim_2}, we have

\begin{equation}
\label{eq:Delta_G_lim_3}
\lVert \Delta \mx G\rVert\leq \omega^2\mu_{\rm 0}\alpha_{\rm e}N_{\rm d}(N_{\rm d}-1)  \left[ \frac{1}{2k_0 a}+\frac{4}{(2k_0 a)^2}+\frac{9}{(2k_0 a)^3}+\frac{9}{(2k_0 a)^4} \right]\frac{k_0(k_0 L_{\rm max})}{\pi}+\omega^2\mu_{\rm 0}\alpha_{\rm e}k_0\mathcal{O}[(k_0L_{\rm max})^{2}].
\end{equation}

\noindent
$\lVert \Delta \mx G\rVert$ can be made arbitrarily small by decreasing $k_0 L_{\rm max}$.

When $k_0 L_{\rm max}\to 0$, we have $\lVert \Delta \mx G\rVert\to 0$ according to \reqn{Delta_G_lim_3}, thus $\lVert \mx F\rVert\to 0$ according to \reqn{F_norm}, thus $\lVert \Delta \ay J^{\rm er} \rVert\to 0$ according to \reqn{J_er_norm}.
The two sides of \reqn{J_cell} become equivalent:

\begin{equation}
\label{eq:J_cell_eq}
    \v J_I = \v J^{\rm inc}_I+\omega^2\mu_{\rm 0}\alpha_{e0} N_I\sum_{J=1,J\neq I}^{N_{\rm c}}\dy G(\v r_{\Omega I}, \v r_{\Omega J})\cdot  \v J_J.
\end{equation}
Notice that $N_I\v J_J\neq 0$ only if there is a particle in cell $\Omega_I$ and another particle in cell $\Omega_J$.
Since we are using the hard sphere model, the distance between the two particles is at least $2a$.
Thus \reqn{J_cell_eq} is equivalent to

\begin{equation}
\label{eq:J_cell_principal}
    \v J_I = \v J^{\rm inc}_I+\omega^2\mu_{\rm 0}\alpha_{e0} N_I\sum_{J=1,\v r_{\Omega J}\notin {\rm PV}_I}^{N_{\rm c}}\dy G(\v r_{\Omega I}, \v r_{\Omega J})\cdot  \v J_J.
\end{equation}
Here ${\rm PV}_I$ is the principal volume used in \reqn{EH} which is a sphere centered at $\v r_{\Omega I}$ with a radius $2a$. We take the expectation values on both sides of \reqn{J_cell_principal}:

\begin{equation}
\begin{split}
\label{eq:J_cell_eq_expectation}
    \langle \v J_I \rangle & = \langle \v J^{\rm inc}_I \rangle+\omega^2\mu_{\rm 0}\alpha_{e0} \sum_{J=1,\v r_{\Omega J}\notin {\rm PV}_I}^{N_{\rm c}}\dy G(\v r_{\Omega I}, \v r_{\Omega J})\cdot \langle  N_I \v J_J \rangle \\
    & = \langle \v J^{\rm inc}_I \rangle+\omega^2\mu_{\rm 0}\alpha_{e0} {\rm P}_I\sum_{J=1,\v r_{\Omega J}\notin {\rm PV}_I}^{N_{\rm c}}\dy G(\v r_{\Omega I}, \v r_{\Omega J})\cdot  \langle \v J_J \rangle_{N_I=1}.
\end{split}
\end{equation}

\noindent
Here $\langle \v J_J \rangle_{N_I=1}$ denotes the expectation value of $\v J_J$ under the condition that there is a dipole in cell $\Omega_I$, i.e. $N_I=1$, and ${\rm P}_I$ is the corresponding probability.

The calculation of the \textit{conditional} expectation $\langle \v J_J \rangle_{N_I=1}$ on rhs of \reqn{J_cell_eq_expectation} requires the knowledge of the probability distribution of the particles in the medium.
Here we apply the diffused particle model in which the probability of finding the $i$th particle at location $\v r$ is given by \reqn{P}.
The initial locations of the particles, $\v r_i$, are preassigned.
So the probability distribution of the particles in the medium is determined by $\sqrt{2D\Delta t}$.
The molecular volume of the medium is given by $V/N$, where $V$ is the volume of the medium.
When $\sqrt{2D\Delta t}\ll \sqrt[3]{V/N}$, only the cells that originally contain particles, are likely to contain particles.
Thus the probability distribution of particles over space is discrete.
At longer time scales, $\sqrt{2D\Delta t}\gg \sqrt[3]{V/N}$, the overall probability density of the particles in the medium is a constant, $N/V$.

In the medium with $\sqrt{2D\Delta t}\ll \sqrt[3]{V/N}$, which is a widely used approximation~\cite{Draine1994,mishchenko_multiple_2006,Mackowski2013DirectParticles}, we can choose such a discretization that the volumetric probability distribution of $i$th particle is entirely included in a cell indexed by $I$.
For this cell, there is always one particle inside the cell, i.e., $N_I=1$, in all configurations.
Thus $\langle \v J_J \rangle_{N_I=1}=\langle \v J_J \rangle$.

The probability density of particles in a medium with $\sqrt{2D\Delta t}\gg \sqrt[3]{V/N}$, i.e. the regime considered in this manuscript, is considered to be a constant.
So we can choose any cell to calculate $\langle \v J_J \rangle_{N_I=1}$ without losing generality.
The cell indexed by $I$ centered at the initial location, $\v r_i$, of $i$th particle, is chosen for the calculation.
Since the number of particles, $N_J$, in cell $J$ only takes the value $1$ or $0$, the conditional expectation of the current in cell $J$ under the condition that ${N_I=1}$ is given by

\begin{equation}
\label{eq:J_exp_cond}
\langle \v J_J \rangle_{N_I=1} = {\rm P}_{J,N_I=1}\langle \v J_J \rangle_{N_I=1,N_J=1}.
\end{equation}
Here ${\rm P}_{J,N_I=1}$ denotes the probability of $N_J=1$ under the condition ${N_I=1}$.
Similarly, for the unconditional expectation, we have

\begin{equation}
\label{eq:J_exp_uncond}
\langle \v J_J \rangle = {\rm P}_J\langle \v J_J \rangle_{N_J=1}.
\end{equation}
The difference between the conditional and unconditional expectation value of the current in cell $J$ is given by

\begin{equation}
\label{eq:J_exp_differ}
\langle \v J_J \rangle_{N_I=1}-\langle \v J_J \rangle = \Delta {\rm P}_J \langle \v J_J \rangle_{N_J=1} + (\langle \v J_J \rangle_{N_I=1,N_J=1} - \langle \v J_J \rangle_{N_J=1}){\rm P}_{J,N_I=1}.
\end{equation}
Here

\begin{equation}
\label{eq:D_P_J}
\Delta {\rm P}_J \equiv {\rm P}_{J,N_I=1} - {\rm P}_J.
\end{equation}
%
If the $i$th particle is in cell $I$, it cannot be in another cell $J$, so we have

\begin{equation}
\label{eq:D_P_J_cal}
\Delta {\rm P}_J = - P(\v r_{\Omega J}, \v r_{\Omega I})V_{J}.
\end{equation}
Here $V_{J}$ is the volume of the cell $J$ and the probability density $P(\v r_{\Omega J}, \v r_{\Omega I})$ can be calculated by \reqn{P}.
Assuming $\sqrt{2D\Delta t}\ll \lambda$, $\langle \v J_J \rangle_{N_J=1}$ is a constant near cell $I$ where $\Delta P_J\neq 0$.
By the symmetry of $P(\v r_{\Omega J}, \v r_{\Omega I})$, we have~\cite{born2013principles,Markel2016IntroductionTutorial}

\begin{equation}
\label{eq:D_P_J_int}
\sum_{J=1,\v r_{\Omega J}\notin {\rm PV}_I}^{N_{\rm c}}\dy G(\v r_{\Omega I}, \v r_{\Omega J})\cdot \Delta {\rm P}_J \langle \v J_J \rangle_{N_J=1}=0.
\end{equation}
%
For the second term in \reqn{J_exp_differ}, the difference between $\langle \v J_J \rangle_{N_I=1,N_J=1}$ and $\langle \v J_J \rangle_{N_J=1}$ is caused by the field scattered from the current in cell $I$ to cell $J$, which is given by

\begin{equation}
\label{eq:single_dipole_scatter}
\langle \v J_J \rangle_{N_I=1,N_J=1} - \langle \v J_J \rangle_{N_J=1} = \omega^2\mu_{\rm 0}\alpha_{e0} \dy G(\v r_{\Omega J}, \v r_{\Omega I})\cdot \langle \v J_I \rangle_{N_I=1}.
\end{equation}
We apply the equation above to \reqn{J_exp_differ}, then we apply \reqn{J_exp_differ} to the $2$nd term in \reqn{J_cell_eq_expectation} to have

\begin{equation}
\label{eq:J_cell_renorm}
    \langle \v J_I \rangle = \langle \v J^{\rm inc}_I \rangle+
    \omega^2\mu_{\rm 0}\alpha_{e0} {\rm P}_I\sum_{J=1,\v r_{\Omega J}\notin {\rm PV}_I}^{N_{\rm c}}\dy G(\v r_{\Omega I}, \v r_{\Omega J})\cdot  \langle \v J_J \rangle+
    \omega^2\mu_{\rm 0}\alpha_{e0} G_{II}\cdot \langle \v J_I \rangle,
\end{equation}
where

\begin{equation}
\begin{split}
\label{eq:G_II}
    G_{II} = \omega^2\mu_{\rm 0}\alpha_{e0} \sum_{J=1,\v r_{\Omega J}\notin {\rm PV}_I}^{N_{\rm c}}\dy G(\v r_{\Omega I}, \v r_{\Omega J})\cdot {\rm P}_{J,N_I=1} \dy G(\v r_{\Omega J}, \v r_{\Omega I}).
\end{split}
\end{equation}
The third term in \reqn{J_cell_renorm} is the current induced in cell $I$ by the dipole fluctuation~\cite{Barrera1988RenormalizedTheory}.
That is, when calculating the $2$nd term in \reqn{J_cell_renorm}, the overall probability density of particles in the medium is considered to be a constant, however, when calculating the $2$nd term in \reqn{J_cell_eq_expectation}, the probability of a dipole existing in cell $I$ is $1$ instead of the averaged probability $N/V$.
This dipole fluctuation can be combined into a renormalized polarizability~\cite{Barrera1988RenormalizedTheory} $\alpha_{e}$, thus \reqn{J_cell_renorm} becomes

\begin{equation}
\label{eq:J_cell_renorm_alpha}
    \langle \v J_I \rangle = \langle \v J^{\rm inc}_{I}  \rangle'+
    \omega^2\mu_{\rm 0}\alpha_e {\rm P}_I\sum_{J=1,\v r_{\Omega J}\notin {\rm PV}_I}^{N_{\rm c}}\dy G(\v r_{\Omega I}, \v r_{\Omega J})\cdot  \langle \v J_J \rangle,
\end{equation}
where
\begin{equation}
\label{eq:renorm_alpha}
    \alpha_e =(I-\omega^2\mu_{\rm 0}\alpha_{e0} G_{II})^{-1}\alpha_{e0}.
\end{equation}
Here $\langle \v J^{\rm inc}_{I}  \rangle'$ in \reqn{J_cell_renorm_alpha} is the current induced in cell $I$ by the incident field but with the renormalized polarizability $\alpha_e$. By \reqn{J_cell_renorm_alpha}, we have shown that the configurational-averaged current can be solved self-consistently.

\appendix
\renewcommand{\theequation}{B\arabic{equation}}
  \setcounter{equation}{0}
\section*{Appendix B: Discretization Of A Half-Space Into Voxels}

\setcounter{equation}{0}

\begin{figure}[b]
\includegraphics{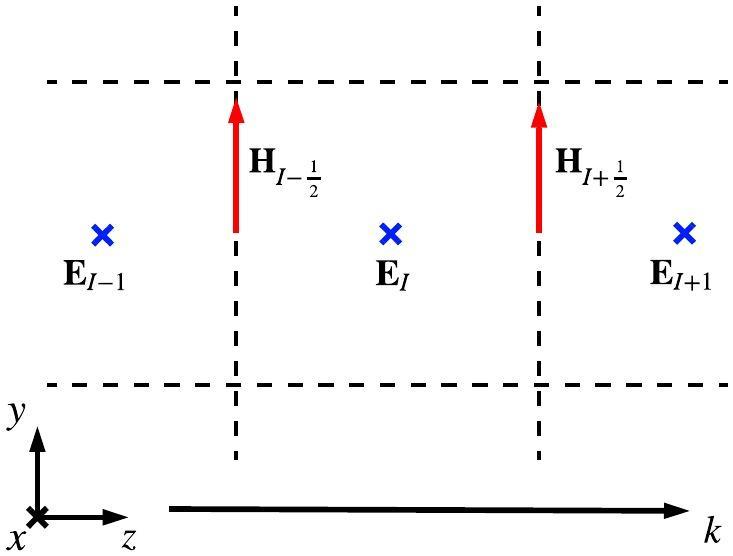}
\caption{\label{fig:Mesh}
The discretization of the simulation region for the electric field calculation. The cross at the voxel center denotes the electric field directed in $x$ axis.}
\end{figure}

In this Appendix, we cluster the particles into voxels in a manner consistent with generalized Foldy-Lax equations presented in the paper.
The whole space is divided into a vacuum half-space and a medium half-space. 
The medium half-space is discretized into voxels, in this paper, each containing 15000 diffused particles, with a voxel size of $0.03\lambda\times 0.03\lambda \times 0.03 \lambda$. 
Yee's lattice~\cite{KaneYee1966NumericalMedia,Correia20043D-FDTD-PMLMetamaterials} is used to separate the electric and magnetic field calculation, given in \rfig{Mesh}. 
This clustering allows us to reduce the Foldy-Lax equation given in \reqn{EH} to a voxel-based equation,

\begin{equation}{\label{eq:EH_d1}}
\begin{split}
\v E_{I}
= 
\v E^{\rm inc}_{I}
+
\sum_{J}\dy G_{IJ}\left[ \omega^2 \mu_0 \alpha_{\rm e1}\v E_{J} + i\omega(\nabla\times \alpha_{\rm m1} \v H)_{J} \right],
\\
\v H_{I}
= 
\v H^{\rm inc}_{I}
+
\sum_{J}\dy G_{IJ}\left[ \omega^2 \varepsilon_0\alpha_{\rm m1} \v H_{J} - i\omega(\nabla\times \alpha_{\rm e1} \v E)_{J} \right].
\end{split}
\end{equation}

\noindent
Here the polarizabilities of the voxel are approximated as $\alpha_{\rm e1} = \rho\Delta V \alpha_{\rm e}$ and $\alpha_{\rm m1} = \rho\Delta V \alpha_{\rm m}$, where $\rho$ is the volume number density of the particles and $\Delta V$ is the volume of the voxel. 
In the limit that the particles form a continuous medium in a half-space, a planewave incident from the free-space side will necessarily produce a planewave in the medium half-space.
This allows us to simplify the equations by imposing an assumption that the field on a medium side is a planewave.
Thus the $\hat{\v y}$ and $\hat{\v z}$ components of all electric fields are omitted as well as the $\hat{\v x}$ and $\hat{\v z}$ components of all magnetic fields, i.e., $\v E \approx \hat{\v x}E$ and $\v H \approx \hat{\v y}H$.
So \reqn{EH_d1} becomes

\begin{subequations}{\label{eq:EH_d2}}
\begin{eqnarray}
\label{eq:EH_d2_1}
\hat{\v x}E_{I}
= 
\hat{\v x}E^{\rm inc}_{I}
+
\sum_{J}\dy G_{IJ}\hat{\v x}\left[ \omega^2 \mu_0\rho \alpha_{\rm e} E_{J} -
i\omega\rho \alpha_{\rm m} (\partial_{\rm z}  H)_{J} \right]\Delta V,
\\
\label{eq:EH_d2_2}
\hat{\v y}H_{I}
= 
\hat{\v y}H^{\rm inc}_{I}
+
\sum_{J}\dy G_{IJ}\hat{\v y}\left[ \omega^2 \varepsilon_0\rho \alpha_{\rm m} H_{J} -i\omega \rho \alpha_{\rm e}( \partial_{\rm z}   E)_{J} \right]\Delta V.
\end{eqnarray}
\end{subequations}

Appealing to Yee's method, we allow $J$ to take on half-integer values to represent a field on the edge of the voxel for the purposes of computing derivatives.
The derivatives of currents are calculated by the central difference method:
\begin{equation}{\label{eq:differ}}
\begin{split}
(\partial_{\rm z}   H)_{J}= & (H_{J+\frac{1}{2}}-H_{J-\frac{1}{2}})/\Delta z,
\\
(\partial_{\rm z}   E)_{J}= & (E_{J+\frac{1}{2}}-E_{J-\frac{1}{2}})/\Delta z.
\end{split}
\end{equation}

As we have assumed, the electric field propagates as a plane wave $E(z) \propto e^{i n k_{\rm 0} z}$ and $H(z) \propto e^{i n k_{\rm 0} z}$, where wavenumber is given by $k_{\rm 0}=\omega\sqrt{\varepsilon_{\rm 0}\mu_{\rm 0}}$ and $n$ is the effective index of refraction of the medium composed of the particles. 
Note that as seen in \rfig{Mesh}, $E_J$ is the field at the center of the voxel, and the voxel is much smaller than the wavelength.
Thus the field on the edge of the voxel is given by
\begin{equation}{\label{eq:edge}}
\begin{split}
E_{J\pm\frac{1}{2}} \approx & E_{J} {\rm exp}(\pm i n k_{\rm 0}\Delta z/2) \approx (1\pm i n k_{\rm 0}\Delta z/2)E_{J},
\\
H_{J\pm\frac{1}{2}} \approx & H_{J} {\rm exp}(\pm i n k_{\rm 0}\Delta z/2) \approx (1\pm i n k_{\rm 0}\Delta z/2)H_{J}.
\end{split}
\end{equation}
Thus the derivatives are found to be
\begin{equation}{\label{eq:differ2}}
\begin{split}
(\partial_{\rm z}   H)_{J} \approx & i n k_{\rm 0}  H_{J},
\\
(\partial_{\rm z}   E)_{J} \approx & i n k_{\rm 0}  E_{J}.
\end{split}
\end{equation}
Plugging the equations above into \reqn{EH_d2} gives discretized equations for the field calculation. 
Further simplification can be achieved by disentangling the electric and magnetic field calculations. 
The magnetic field can be eliminated from \reqn{EH_d2_1} by making use of fact that $H = E/(\eta\eta_{\rm 0})$, and similarly the electric field may be eliminated from \reqn{EH_d2_2} by noting that $E = \eta\eta_{\rm 0} H$, where $\eta\eta_{\rm 0}$ is the wave impedance of the propagation field.

\appendix
\renewcommand{\theequation}{C\arabic{equation}}
  \setcounter{equation}{0}
\section*{Appendix C: The Precalculation of $d_{\rm F}$}

In this Appendix, we describe the strategy to find the value of $d_{\rm F}$ to balance the accuracy and running time of the algorithm.
Recall that $d_{\rm F}$ is the distance scale that separates elements that are in the far-field of each other and can thus be clustered together at the lv1-lv2-lv3 clustering scheme from elements that are in the near field of each other and must be treated with the lv1-lv3 clustering scheme.
The lv1-lv2-lv3 scheme is less computationally expensive, and so we choose it when we can.
Discretizing the lv3 column by a coarse mesh by introducing the intermediate lv2 voxel reduces computational complexity.
However, the lv0-lv1-lv2-lv3 clustering is reliable only if the difference between the fields scattered by all the lv1 voxels and by all the lv2 voxels in the same lv3 column can be omitted.

We compare the electric field $\v E^{\rm sca}_{\rm 1}$ scattered from a column composed by the lv1-lv2-lv3 clustering illustrated in \rfig{lv2}(a) with $\v E^{\rm sca}_{\rm 2}$ from a column composed by the lv1-lv3 clustering illustrated in \rfig{lv2}(b).
The error is defined by the relative difference between these two scattered fields along the $z$ axis, given by
\begin{equation}
\label{eq:error2}
    e_{\rm col} = \frac{| \v E^{\rm sca}_{\rm 2} - \v E^{\rm sca}_{\rm 1}|}{|\v E^{\rm sca}_{\rm 1}|}.
\end{equation}

\noindent
Both fields, $\v E^{\rm sca}_{\rm 1}$ and $\v E^{\rm sca}_{\rm 2}$, are calculated by the 2nd term in \reqn{EH_d3}. 
For the purpose of simplicity, only the $x$ components of the electric fields are considered because the $y$ and $z$ components vanish. 
The calculated error is plotted in \rfig{lv2}(c). 
It can be observed that the difference between the two clustering methods falls to negligible for distances between the columns $r$ such that $r>0.1\lambda$. 
For $r<0.1\lambda$, however, the error of applying the lv2 voxel in the hierarchical clustering process can not be neglected thus only lv1-lv3 clustering can be chosen when calculating $M_{\mathbb{I}\mathbb{J}}$. 
Thus the condition distance in \rfig{preprocessing} denoting the limit between far field and near field, $d_{\rm F}$, is taken to be $0.1\lambda$.

\begin{figure}
\includegraphics{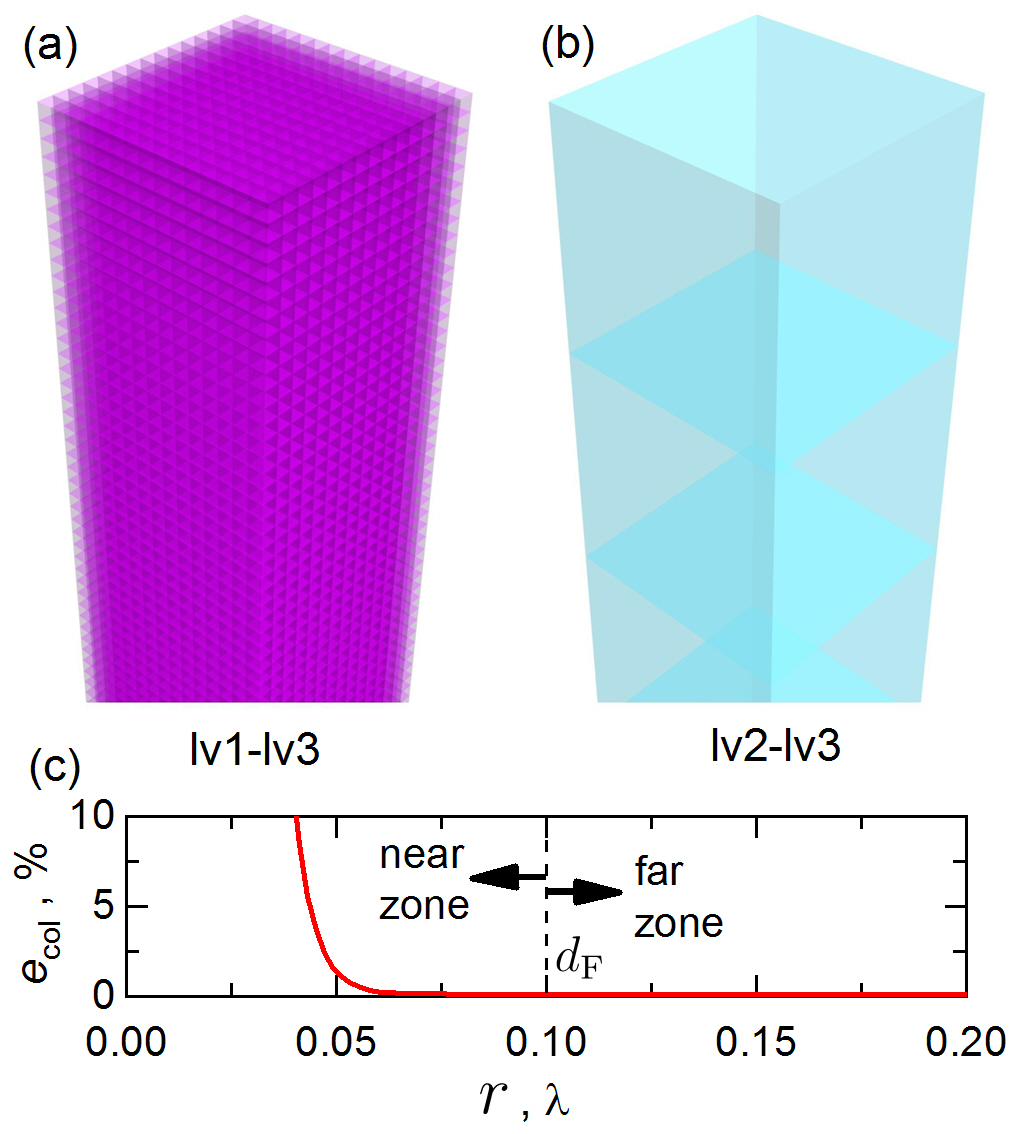}
\caption{\label{fig:lv2}
(a) A lv3 column composed of lv1 voxels. (b) A lv3 column composed of lv2 voxels. The electric fields scattered by both structures are calculated along the straight line perpendicular to the column on the $yz$ plane, starting from the surface of the column. 
(c) The relative difference between the scattered fields.}
\end{figure}

\appendix
\renewcommand{\theequation}{D\arabic{equation}}
  \setcounter{equation}{0}
\section*{Appendix D: Validity Of The Hierarchical Clustering}

\setcounter{equation}{0}

In this Appendix, we check the validity of the hierarchical clustering procedure.
We consider a planewave with a wave vector $\hat{\v z}n k_{\rm 0}$ propagating in a lv2 cube and calculate the far field scattered by the cube, illustrated in \rfig{lv012er}(a), where the refractive index $n$ is acquired after the convergence of the main algorithm.
The hierarchical clustering process is reliable if the fields in the far zone scattered by the clustered structures at levels 0, 1 and 2 are all approximately equal, that is if $\v E^{\rm sca}_{\rm 0} \approx \v E^{\rm sca}_{\rm 1} \approx \v E^{\rm sca}_{\rm 2}$.
The locations of the diffused particles (i.e. lv0 voxels) are generated randomly inside and near the lv2 box.
The variance of particle diffusion $2D\Delta t$ is taken to be $6\times 10^{-4} \lambda^2$, \footnote{The value is approximated (with one significant digit) with a diffusivity $D= 2\times 10^{-5}{\rm cm^2/s}$ corresponding to the Brownian motion of air molecules dissolved in water and a diffusion time $\Delta t = 4$ ns which is enough for the electromagnetic fields to reach to a stable distribution in the simulated medium with a wavelength of $221$ nm.}.
The particles located at the distance more than $0.15\lambda$ outside of the lv2 box are ignored \footnote{much larger than the standard deviation of the Gaussian distribution of the particle diffusion $\sqrt{2D\Delta t} = 0.0245\lambda$}.
The calculation of $\v E^{\rm sca}_{\rm 0}$ is given by the 2nd term in \reqn{EH_1}.
The detailed calculation method including how to deal with the curl of the currents is given in Appendix B.
The fields scattered by the lv1 and lv2 structures, $\v E^{\rm sca}_{\rm 1}$ and $\v E^{\rm sca}_{\rm 2}$, are given as the second term on the right-hand side of \reqn{EH_d3_1}, where the parameters $n$ and $\eta$ are given by the simulation results. 

\begin{figure}[b!]
\includegraphics{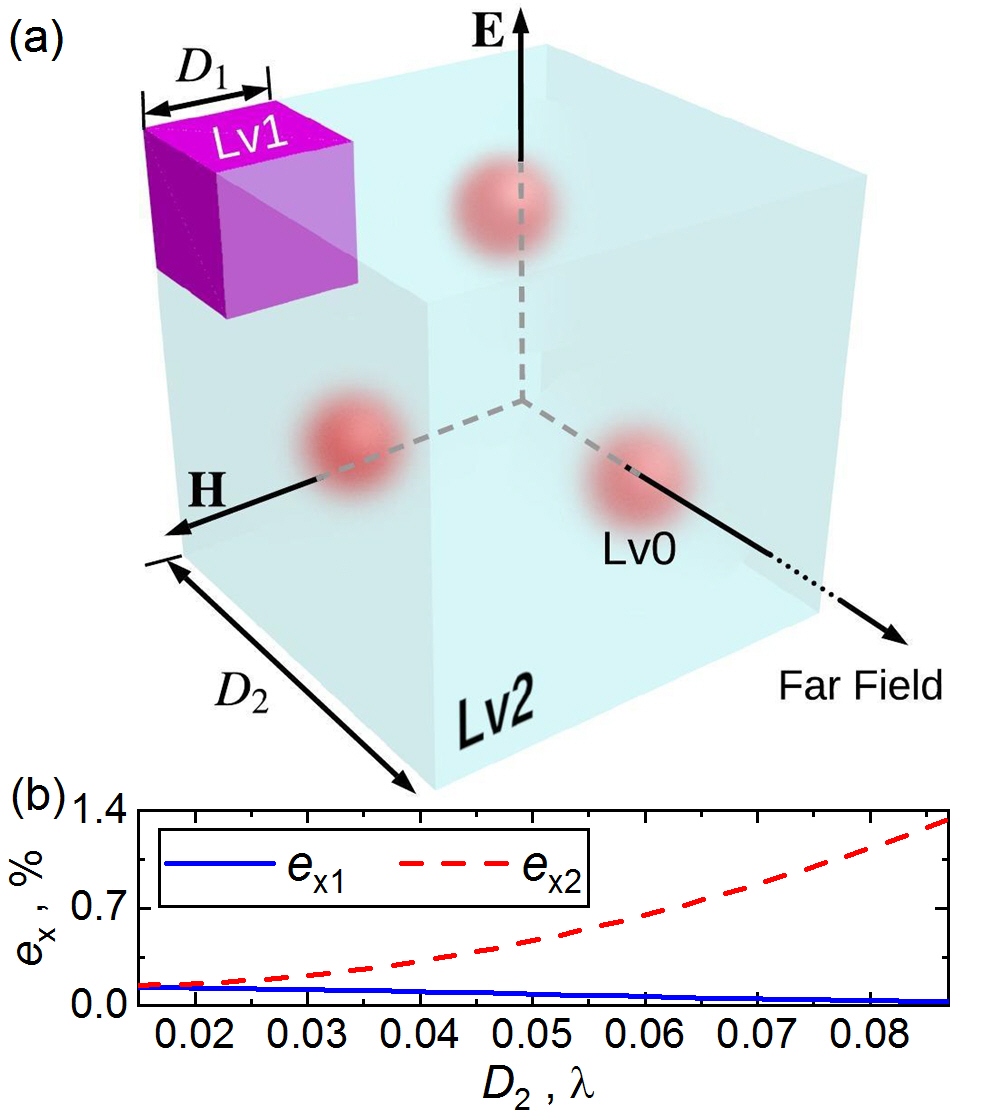}
\caption{\label{fig:lv012er}
(a) The model of simulating the fields scattered by lv0, lv1 and lv2 structures and (b) the errors between $\v E^{\rm sca}_{\rm 0}$ and $\v E^{\rm sca}_{\rm 1}$ and between $\v E^{\rm sca}_{\rm 0}$ and $\v E^{\rm sca}_{\rm 2}$ with different $\Delta L_{\rm 2}$ values. The scattered fields are calculated in the far field region.}
\end{figure}

The relative difference between the $x$ component of the electric fields scattered from the lv0 structures and from the lv1 or lv2 structures are defined as

\begin{equation}
\label{eq:error}
    e_{\rm x1,x2} = \frac{| \v E^{\rm sca}_{\rm x1,x2} - \v E^{\rm sca}_{\rm x0}|}{|\v E^{\rm sca}_{\rm x0}|}.
\end{equation}

\noindent
The dependence of these clustering errors on $D_2$, the size of the lv2 voxel, is shown in \rfig{lv012er}(b).
It can be observed that taking the lv2 side length, $\Delta L_{\rm 2}$, to be $0.03\lambda$ limits the clustering error on the scattered electric field to be smaller than $0.3\%$, which results in a maximum error of $0.6\%$ in the simulated refractive index according to \reqn{n}.
In applications where the error of refractive index calculation is required to be lower than $0.6\%$, a smaller sized lv2 cube should be chosen based on \rfig{lv012er}(b) for the control of accuracy.

\end{widetext}


%

\end{document}